\documentclass[graybox]{svmult}

\usepackage{mathptmx}       
\usepackage{helvet}         
\usepackage{courier}        
\usepackage{type1cm}        
\usepackage{makeidx}         
\usepackage{graphicx}        
\usepackage{natbib} 
\usepackage{multicol}        
\usepackage[bottom]{footmisc}
\usepackage{hyperref}

\makeindex             

\def\aap{A\&A}

\def\apj{Astrophys. Journ.}
\def\aj{Astron. Journ.}
\def\apjl{Astrophys. Journ. Lett.}

\def\mnras{Month. Not. Roy. Ast. Soc.}
\def\pasp{Pub. Ast. Soc. Pac.}
\def\iaucirc{IAU circ.}

\def\kms{\,${\rm km}\cdot{\rm s}^{-1}$\,}

\begin{document}

\title*{Interactions in massive binary stars\index{star~(binary)} as seen by interferometry\index{interferometry}}
\author{F. Millour, A. Meilland, P. Stee, O. Chesneau}
\institute{F. Millour \at OCA, Bd de l'Observatoire, 06304 Nice, \email{fmillour@oca.eu}
\and A. Meilland \at OCA, Bd de l'Observatoire, 06304 Nice,  \email{ame@oca.eu}}
%
%
\maketitle

\abstract{With the advent of large-collecting-area instruments, the
  number of objects that can be reached by optical long-baseline
  interferometry\index{interferometry} is steadily increasing.  We
  present here a few results on massive binary
  stars\index{star~(binary)}, showing the interest of using this
  technique for studying the insight of interactions in these
  systems. Indeed, many massive stars\index{star~(massive)} with
  extended environments host, or are suspected to host, companion
  stars. These companions could have an important role in shaping the
  circumstellar environment\index{circumstellar~environment} of the
  system. These examples provide a view in which
  binarity\index{star~(binary)} could be an ingredient, among many
  others, for the activity of these stars\index{star~(active)}.}

\section{Introduction}
\label{sec:1}

The most massive stars\index{star~(massive)} are still a puzzle,
especially regarding their evolution. They are the best candidates to
be progenitor of type II supernovas\index{supernova}, which are one
source of metallic enrichment of the interstellar medium, but they
also input kinetic energy to their vicinity, triggering densification
and collapse of neighboring interstellar clouds.

In the past years, many new observing techniques have appeared, that
allow one to access unprecedented angular resolution. Therefore, the
detection and characterization of circumstellar
environments\index{circumstellar~environment} (hereafter CSEs) around massive
stars\index{star~(massive)} has become a reality. A by-product is that
several new companion stars have been discovered using high-angular
resolution techniques, being adaptive optics, lucky imaging, or
long-baseline interferometry\index{interferometry}.

We try here to give an insight of the influence of stellar companions
to the structure of CSEs\index{circumstellar~environment} by showing a
few examples of massive stars\index{star~(massive)}, ranging from the
classical Be stars\index{star~(Be)} up to the most massive blue
supergiant stars\index{star~(supergiant)}.

\section{Interactions in intermediate-mass stars: classical Be stars\index{star~(Be)}}

Binarity\index{star~(binary)} may play a non-negligible role in the
formation of circumstellar disks\index{circumstellar~disk} around
classical Be stars\index{star~(Be)}. These stars close to main
sequence are known to be fast rotators, however the most recent
studies \citep{2005A&A...440..305F, 2005ApJ...634..585C, 2011Be}
indicate that they are not all critical-rotators, so rotation cannot
be the only mechanism involved in the mass-ejection. Recent studies
like \cite{2011A&A...528A..48M} try to investigate the role of the
gravitational influence of a companion star in a few extreme
cases. However, such influence is not the only physical process
proposed as an additional momentum source to cancel the stellar
gravity. Radiative pressure \citep{1979IAUS...83..237A}, non-radial
pulsations \citep{1998A&A...336..177R}, or even magnetism
\cite{2008ApJ...672.1174L} have also been suggested to play such a
role.

To progress in the understanding of the physical processes responsible
for the Be phenomenon one needs to resolve their
CSE\index{circumstellar~environment} both spatially to obtain
informations on the distribution of matter surrounding the star, but
also spectrally to get access to the kinematics of the ejected
matter. Consequently, spectro-interferometry\index{interferometry} is
the most suitable technique to study these objects. First
VLTI\index{VLTI}/AMBER\index{AMBER} \citep{2007A&A...464....1P} and
CHARA\index{CHARA}/VEGA\index{VEGA} \citep{2009A&A...508.1073M}
observations of Be stars\index{star~(Be)} have evidenced that the
matter is mostly concentrated into the equatorial plane and that this
geometrically thin disk is dominated by rotation with a rotational law
close to the Keplerian\index{Kepler} one \citep{2007A&A...464...59M,
  2007A&A...464...73M, 2011Be, 2009A&A...504..915C,
  2011A&A...529A..87D, 2011arXiv1109.3447K}.

\begin{figure*}[htbp]
       \centering  
       \includegraphics[width=0.95\textwidth]{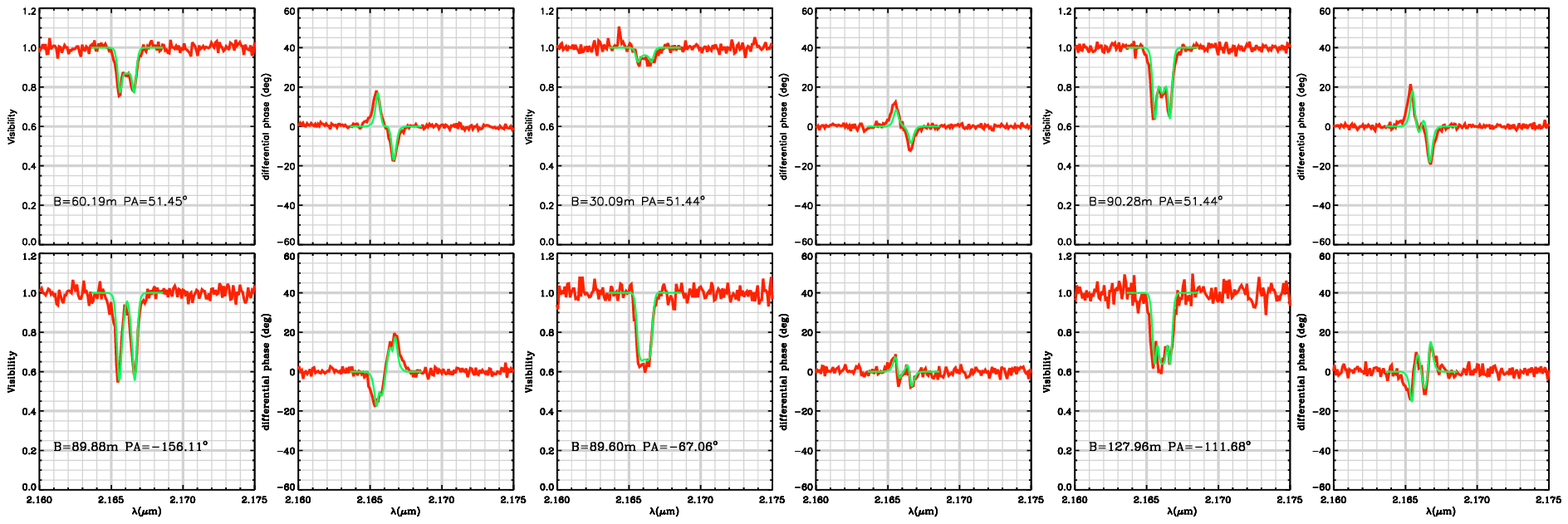}
\caption{$\alpha$~Col\index{$\alpha$~Col} selected differential
  visibilities\index{visibility~(interferometry)} and
  phases\index{phase~(interferometry)} from 2
  VLTI\index{VLTI}/AMBER\index{AMBER} HR measurements (red line). Each
  row corresponds to one VLTI\index{VLTI}/AMBER\index{AMBER}
  measurement (3 different baselines). The top row shows short
  baselines (barely resolved disk) while the bottom row shows long
  baselines (fully resolved disk). The
  visibilities\index{visibility~(interferometry)} and
  phases\index{phase~(interferometry)} of the best-fit
  geometrically-thin-rotating-disk model is over-plotted in
  green. From \citet{2011Be}.}
\label{alpcol}
\end{figure*}

An example of spectro-interferometric observations with the
VLTI\index{VLTI}/AMBER\index{AMBER} well-fitted by a simple
geometrically thin rotating disk model is plotted in
Fig~\ref{alpcol}. As seen in this figure, a rotating disk provides
typical "W"-shaped visibilities\index{visibility~(interferometry)} and
"S"-shaped phases\index{phase~(interferometry)} as long as the disk is
not fully resolved by the interferometer (top plots), whereas it gives
more complex phases\index{phase~(interferometry)} shape (double-"S")
when it is fully resolved (bottom plots).

\begin{figure*}[htbp]
       \centering  
       \includegraphics[width=0.5\textwidth]{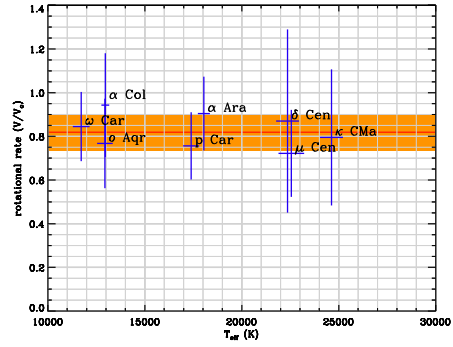}
\caption{A set of 8 Be stars\index{star~(Be)} observed with long-baseline
  interferometry\index{interferometry}, showing for the first time the
  rotation rate of the central star free of any inclination effect. We
  show here that the rotation rate is roughly constant whatever the
  star temperature, at $C\approx80\%$ of the critical rotation. This
  points to a different process than rapid rotation for all the stars
  to expel their circumstellar disks\index{circumstellar~disk}. From
  \citet{2011Be}.}
\label{BeVcrit}
\end{figure*}

Few interferometric observations also suggest the presence of a
non-negligible polar wind \citep{2006A&A...453.1059K,
  2007A&A...464...59M}. Moreover, thanks to interferometric
measurements of the projected disk flattening, the star inclination
angle can be inferred without ambiguities, and consequently, the
rotational velocity of the central star can be determined. In the
first spectro-interferometric survey of Be star\index{star~(Be)},
\citet{2011Be} show that the stars are rotating at 82$\pm$0.07$\%$ of
their critical velocity (V$_c$), a result compatible with estimation
from \citet[][see Fig.~\ref{BeVcrit}]{2005A&A...440..305F}. Finally,
interferometry\index{interferometry} can also help to discover new
companions as evidenced in the case of $\delta$~Cen\index{$\delta$~Cen} \citet{2008A&A...488L..67M}.

Some Be stars\index{star~(Be)} show large
variability\index{star~(active)} that may be related to
binarity\index{star~(binary)}. For instance, this is the case of two
well studied objects : Achernar\index{Achernar} and
$\delta$~Sco\index{$\delta$~Sco}. Achernar\index{Achernar} is a
quasi-cyclic Be star\index{star~(Be)} with a period of formation and
dissipation of the equatorial disk of about 12 years. It was observed
at a minimum of activity by \citet{2003A&A...407L..47D} and the
authors found that the star was strongly flattened by
close-to-critical rotation. Using these interferometric observations
as well as spectroscopic follow-up of a full cycle of activity from
\citet{2006A&A...446..643V, 2008A&A...486..785K} managed to fully
model the environment of this object which consists in a steady polar
wind driven by radiative pressure and a transient equatorial disk
which is produced during a brief outburst. The ejected matter then
propagates into the CSE\index{circumstellar~environment} with an
expansion velocity of the order of 0.2\kms. The cause of the ejection
remained mysterious until the discovery of a companion star using the
VLT/NACO instrument \citep{2008A&A...484L..13K}. New observations will
soon be executed to constrain the companion orbit and determine
whether or not it could be the cause of the cyclic variations of
Achernar\index{Achernar}.

\begin{figure*}[htbp]
       \centering  
       \includegraphics[width=0.95\textwidth]{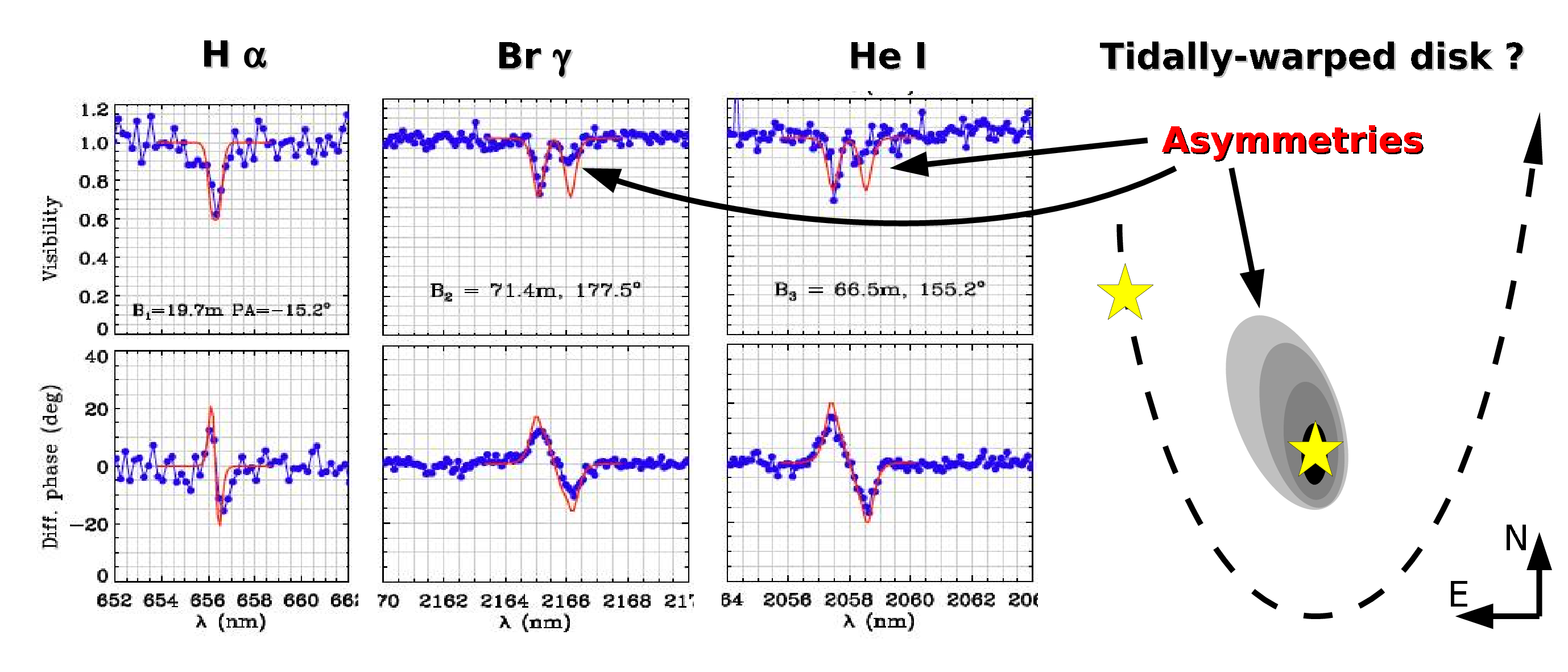}
\caption{Illustration of the $\delta$~Sco\index{$\delta$~Sco} system
  close to the periastron passage. The three left figures are from
  \citet{2011A&A...532A..80M}, showing the AMBER\index{AMBER} data
  (blue) together with a kinematic model of a rotating disk
  (red). \citet{2011A&A...532A..80M} proposed that the differences
  between the model and the data (arrows) could be the signature of a
  tidally-warped disk around the primary star (right sketch).}
\label{deltasco}
\end{figure*}

On the other hand, $\delta$~Sco\index{$\delta$~Sco} is a well-known
binary system\index{star~(binary)}. First evidence of its multiplicity
was reported by \citet{1901MNRAS..61..358I} using the
lunar~occultation\index{lunar~occultation} technique. However, this
work was forgotten for a long time, and the
binary\index{star~(binary)} nature of $\delta$~Sco\index{$\delta$~Sco}
was rediscovered with three different techniques in 1974: by
speckle-interferometry\index{interferometry}
\citep{1974ApJ...194L.147L},
lunar~occultation\index{lunar~occultation} \citep{Dunham1974}, and
intensity interferometry\index{interferometry}
\citep{1974MNRAS.167..121H}. The extremely eccentric orbit
(e$\simeq$0.94) was then constrained by many authors using speckle and
long-baseline interferometry\index{interferometry}
\citep{1993AJ....106..768B, 1996AJ....111..370H, 2001A&A...377..485M,
  2009MNRAS.396..842T, 2011ApJ...729L...5T}. However, the
$\delta$~Sco\index{$\delta$~Sco} system did not show clear evidence of
the Be phenomenon until its June 2000 periastron. At this epoch,
\citet{2001IBVS.5026....1O} found a 0.4\,mag brightening of the
object. Simultaneous spectroscopic observations published in
\citet{2000IAUC.7461....1F} showed evidences of strong
H$\alpha$\index{H$\alpha$} emission lines. Using spectroscopic
measurements obtained at different epochs and assuming that the matter
was concentrated in a Keplerian\index{Kepler} rotating disk,
\cite{2003A&A...408..305M} suggested that the ejected matter was
expanding at about 0.4\kms, a velocity compatible with the one found
for Achernar\index{Achernar}. Thanks to new spectrally-resolved
VLTI\index{VLTI}/AMBER\index{AMBER} and
CHARA\index{CHARA}/VEGA\index{VEGA} observations,
\citet{2011A&A...532A..80M} managed to probe the
CSE\index{circumstellar~environment} structure and confirmed
Miroshnichenko hypothesis that the ejected matter is concentrated in a
Keplerian\index{Kepler} rotating disk. They also constrained the
stellar rotational velocity and found that this star was likely to
rotate far below its critical limit, at about 0.7V$_c$. Finally, they
detect asymmetries in their data that can hardly be modeled under
\cite{1997A&A...318..548O} one-armed oscillation framework, and that
could be due to a tidal warping of the disk by the companion
periastron passage (see Fig.\ref{deltasco}), as is described in
\cite{2011A&A...528A..48M}.

In the case of these two stars, the binarity\index{star~(binary)}
clearly play a important role in the mass-ejection process. However,
whether the companion action is direct, i.e. by canceling the residual
stellar gravity at the surface of the central star, or indirect,
i.e. by impacting on other physical processes such as non-radial
pulsations, remains an open question. Moreover, the influence of
binarity\index{star~(binary)} on the Be phenomenon is not limited to the
mass-ejection, it can also affect the reorganization of the gas in the
CSE\index{circumstellar~environment}.

\section{Interactions in the most massive stars\index{star~(massive)}}

As seen in the previous section, disks encountered around Be
stars\index{star~(Be)} are more and more found to be rotating close to
Keplerian\index{Kepler} velocities. This applies also to some
B[e]\index{star~(B[e])} candidate supergiant
stars\index{star~(supergiant)}.

The situation for Be stars\index{star~(Be)} is somewhat complicated,
as several hypotheses exist to explain the disk formation and
steadiness. For B[e] stars\index{star~(B[e])}, which are surrounded by
dense disks of plasma {\it and} dust, the situation is even more
complex. The dust survives much closer to these hot star than expected
so far \citep{2010A&A...512A..73M, 2009A&A...507..317M,
  2007A&A...464...81D}, meaning that complex radiative transfer
processes such as line-blanketing could occur in the gas disk of these
stars. The B[e]\index{star~(B[e])} supergiant
stars\index{star~(supergiant)} critical rotation rate is strongly
decreased by the increase of their radius while leaving the main
sequence. Therefore, rotation alone is certainly not sufficient to
explain the creation of a circumstellar disk\index{circumstellar~disk}
without invoking the influence of a close companion
\citep{2007ApJ...667..497M}.

\begin{figure*}[htbp]
       \centering  
       \includegraphics[width=0.95\textwidth]{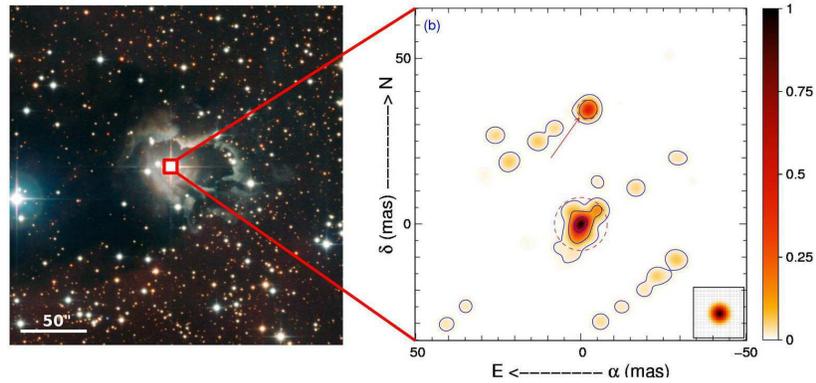}
\caption{Images of HD87643\index{HD\,87643} at two highly (1:1000)
  different scales. Left is the large-scale nebula, exhibiting its
  complex shape and its inner arc-like structures, right is the
  interferometric small-scale image, showing the companion star
  (arrow) and the circumstellar disk\index{circumstellar~disk} (dotted
  circle). Figures from \citet{2009A&A...507..317M}.}
\label{HD87643}
\end{figure*}

Such influence seems clear in the few cases where companion stars were
detected, like in the binary\index{star~(binary)} system
HD\,87643\index{HD\,87643} \citep[][see
  Fig.~\ref{HD87643}]{2009A&A...507..317M}. In this B[e]
system\index{star~(B[e])}, enshrouded inside a complex nebula
reminiscent of the ones found around LBVs, the presence of the
companion provides a key to understand the whole range of features
observed in the system, if it has an eccentric orbit:
\begin{itemize}
\item the extended nebula, that could come from a past outburst of the
  system at one periastron passage,
\item the series of arc-like structures found in the same nebula,
  tracing more recent periastron passages
\item the main star disk, formed by the direct interaction with the
  companion star
\end{itemize}

\begin{figure*}[htbp]
       \centering  
       \includegraphics[width=0.95\textwidth]{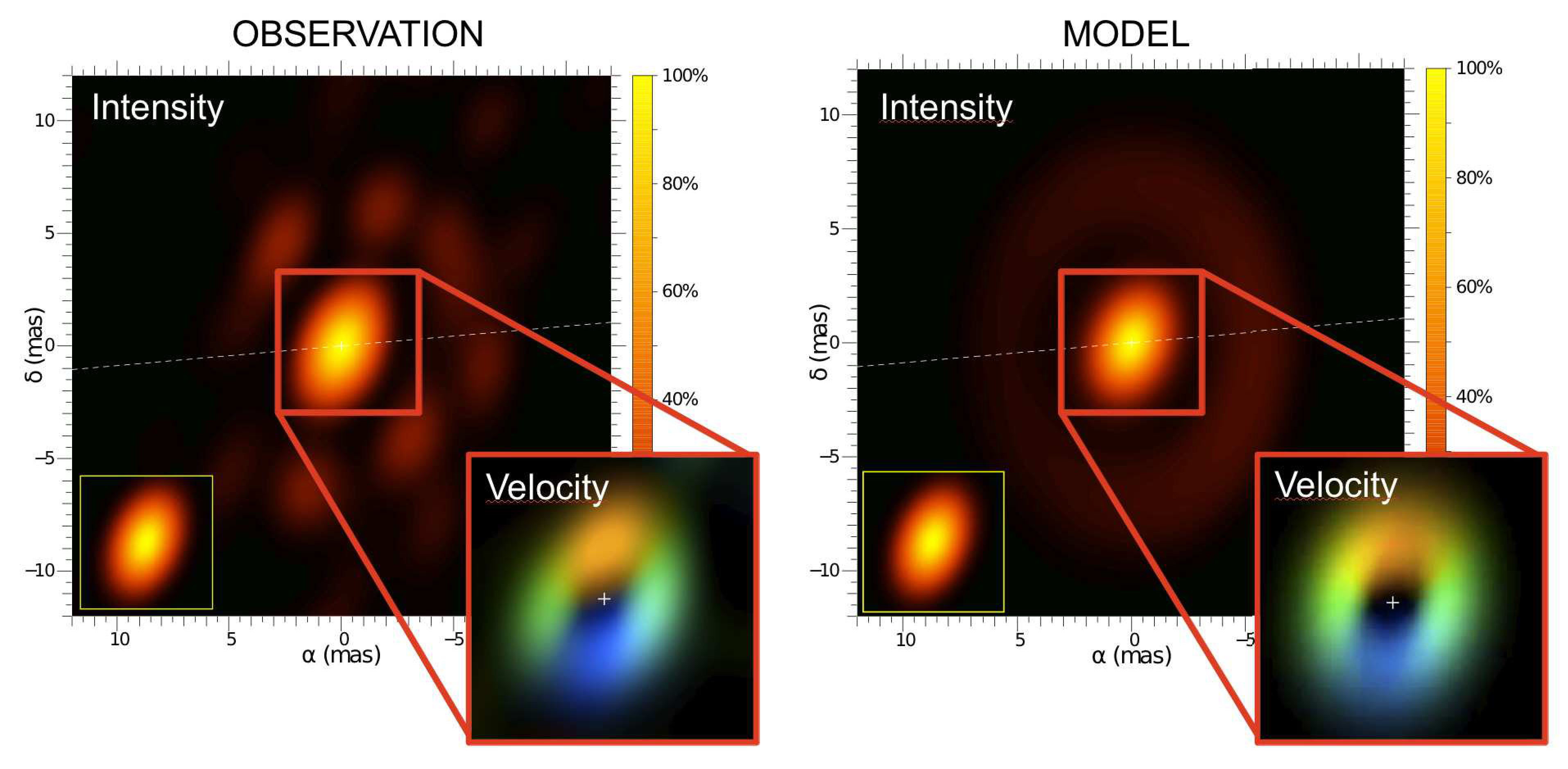}
\caption{The VLTI\index{VLTI} observations of the A[e] system
  HD62623\index{HD\,62623} (here, the figure of the
  \href{http://www.mpifr-bonn.mpg.de/public/pr/pr-hd62623-en.html}{press
    release}) showed that the dense circumbinary~disk\index{circumbinary~disk} \citep[][outer red ring in both the
    image -- left -- and the model -- right]{2010A&A...512A..73M}
  seems constantly fed by an inner plasma disk (central yellow dot),
  probably in Keplerian\index{Kepler} rotation (insets), whose angular
  momentum originates from a solar mass companion
  \citep{2011A&A...526A.107M}.}
\label{HD62623}
\end{figure*}

The supergiant\index{star~(supergiant)} A[e] star
HD\,62623\index{HD\,62623} is also informative in this context
\citep{2011A&A...526A.107M, 2010A&A...512A..73M}. It is an A
supergiant\index{star~(supergiant)} showing the ``B[e]
phenomenon''\index{star~(B[e])}, namely a spectrum dominated by strong
emission lines (allowed and forbidden), and a large infrared
excess. Spectrally and spatially resolved
AMBER\index{AMBER}/VLTI\index{VLTI} observations in the
Br$\gamma$\index{Br$\gamma$} line have shown that the supergiant
star\index{star~(supergiant)} lies in a cavity, and is surrounded by a
rotating disk of plasma. The Br$\gamma$\index{Br$\gamma$} line at the
location of the central star is {\it in absorption} showing that it is
a normal A-type star, albeit with a significantly large v$sini$ of
about 50\kms. By contrast, the Balmer\index{Balmer~line} and
Bracket~lines\index{Brackett~line} are wider ($vsini$ of about
120\kms), and the AMBER\index{AMBER} observations demonstrated that
they originate from a disk of plasma, most probably in
Keplerian\index{Kepler} rotation (See Fig.~\ref{HD62623}). In absence
of any proof of binarity\index{star~(binary)}, it is often difficult
to understand how such a dense equatorial disk could have been
generated. However, HD62623\index{HD\,62623} is a known
binary\index{star~(binary)} with a stellar companion that orbits close
to the supergiant\index{star~(supergiant)} with a period of about 136
days \citep{1995A&A...293..363P}. The mass ratio inferred is very
large, and the companion is probably a solar-mass star, hence unseen
in the AMBER\index{AMBER} images. \citet{1995A&A...293..363P} proposed
that an efficient angular momentum transfer occurs near the L2
Lagrangian point of the system, propelling the mass lost from the
supergiant\index{star~(supergiant)} by its radiative wind and probably
also by strong tides into a stable dense
circumbinary~disk\index{circumbinary~disk}.

Similarly, the evolved system $\upsilon$\,Sgr\index{$\upsilon$\,Sgr}
was recently investigated-again using spectroscopy and optical
interferometry\index{interferometry} in the near-IR and the visible
\citep{2011A&A...532A.148B, 2009A&A...499..827N, 2006A&A...459..849K},
evidencing a dense circumbinary~disk\index{circumbinary~disk},
apparently long-lived.

\section{Novas\index{nova}, bipolar nebulae, and the underlying binary system\index{star~(binary)}}

Novas\index{nova} are formed of a white dwarf (WD) and a red giant
star, whose material is accreted on the WD surface. When enough
material is accreted, thermonuclear reactions can light up, expelling
a fireball, which we directly observe, usually using spectroscopy, or
imaging a few years after. Novas\index{nova} are suspected to be the
progenitors of type I supernovas\index{supernova}, as the accreted
material piles-up, nova\index{nova} after nova, providing the good
conditions for core-collapse.

Such fireballs are seen as bipolar nebulae years
after the explosion, but an open question remained to know if that was
an intrinsic shape of the explosion or if it was shaped after, by the
CSE\index{circumstellar~environment}.

A new field of research was open by the first spectro-interferometric
observations of a nova\index{nova} in
\citet{2007A&A...473L..29C}. This letter evidenced an elongation of
the nova\index{nova} fireball, just a gasp after the outburst (5.5
days). This first clue of bipolarity for one nova\index{nova} could not
be repeatedly achieved when observing a second nova\index{nova} with
the VLTI\index{VLTI} \citep{2008A&A...487..223C}, due to a different
observing configuration. Nevertheless, more recently, \citet{2011TPyx}
provided firm evidence of bipolarity on a third nova\index{nova},
observed in 2011.

The presence of bipolar nebulae at the very first moments of outbursts
in massive stars\index{star~(massive)} (e.g. LBVs like
$\eta$~Car\index{$\eta$~Car}) or less massive systems (e.g. planetary
nebulae, novas\index{nova}) is still a broad and controversial
subject. Indeed, bipolarity in the nebula is very often associated
with binarity\index{star~(binary)} in the core
\citep{2009PASP..121..316D}. But the inverse is not true, namely that
binarity\index{star~(binary)} will essentially imply at a moment of
the life of the system a bipolar ejection of material.

The key question is therefore to link the initial parameters of the
system to:
\begin{itemize}
\item its evolution,
\item the formation of a circumstellar disk\index{circumstellar~disk}
  (such as the ones encountered around B[e]\index{star~(B[e])} stars
  or interacting systems such as
  $\upsilon$\,Sgr\index{$\upsilon$\,Sgr}, or
  $\delta$~Sco\index{$\delta$~Sco}),
\item the occurrence of outburst events, forming rapidly a dense
  bipolar nebula.
\end{itemize}

but it still has to be answered.

\section{Discussion, concluding remarks}

With this paper, we tried to link the presence of a companion star to
the presence of CSEs\index{circumstellar~environment}, by providing
recent examples of high-angular resolution observations.

interferometry\index{interferometry} has brought the view of highly
structured CSEs\index{circumstellar~environment} around several
massive stars\index{star~(massive)}, in the form of thin or thick
disks, plus sometimes the presence of a companion star. The formation
scenario of these CSEs\index{circumstellar~environment} remains to be
determined, but the ingredients seem to count, among others, a binary
system\index{star~(binary)}, including a hot, massive
star\index{star~(massive)}, a dense disk, and perhaps also a
fast-rotating star.

\begin{acknowledgement}
The authors are thankful to the organizers of this very nice
school. We also thank the VLTI\index{VLTI} team which is improving
this wonderful instrument at every observation.
\end{acknowledgement}

\printindex


\begin{thebibliography}{47}
\expandafter\ifx\csname natexlab\endcsname\relax\def\natexlab#1{#1}\fi

\bibitem[{{Abbott}(1979)}]{1979IAUS...83..237A}
{Abbott}, D.~C. 1979, in IAU Symposium, Vol.~83, Mass Loss and Evolution of
  O-Type Stars, ed. {P.~S.~Conti \& C.~W.~H.~De Loore}, 237--239

\bibitem[{{Bedding}(1993)}]{1993AJ....106..768B}
{Bedding}, T.~R. 1993, \aj, 106, 768

\bibitem[{{Bonneau} {et~al.}(2011){Bonneau}, {Chesneau}, {Mourard},
  {B{\'e}rio}, {Clausse}, {Delaa}, {Marcotto}, {Perraut}, {Roussel}, {Spang},
  {Stee}, {Tallon-Bosc}, {McAlister}, {Ten Brummelaar}, {Sturmann}, {Sturmann},
  {Turner}, {Farrington}, \& {Goldfinger}}]{2011A&A...532A.148B}
{Bonneau}, D., {Chesneau}, O., {Mourard}, D., {et~al.} 2011, \aap, 532, A148+

\bibitem[{{Carciofi} {et~al.}(2009){Carciofi}, {Okazaki}, {Le Bouquin}, {{\v
  S}tefl}, {Rivinius}, {Baade}, {Bjorkman}, \& {Hummel}}]{2009A&A...504..915C}
{Carciofi}, A.~C., {Okazaki}, A.~T., {Le Bouquin}, J.-B., {et~al.} 2009, \aap,
  504, 915

\bibitem[{{Chesneau} {et~al.}(2008){Chesneau}, {Banerjee}, {Millour},
  {Nardetto}, {Sacuto}, {Spang}, {Wittkowski}, {Ashok}, {Das}, {Hummel},
  {Kraus}, {Lagadec}, {Morel}, {Petr-Gotzens}, {Rantakyro}, \&
  {Sch{\"o}ller}}]{2008A&A...487..223C}
{Chesneau}, O., {Banerjee}, D.~P.~K., {Millour}, F., {et~al.} 2008, \aap, 487,
  223

\bibitem[{{Chesneau} {et~al.}(2007){Chesneau}, {Lykou}, {Balick}, {Lagadec},
  {Matsuura}, {Smith}, {Spang}, {Wolf}, \& {Zijlstra}}]{2007A&A...473L..29C}
{Chesneau}, O., {Lykou}, F., {Balick}, B., {et~al.} 2007, \aap, 473, L29

\bibitem[{{Chesneau} {et~al.}(2011){Chesneau}, {Meilland}, {Banerjee}, {Le
  Bouquin}, {toto}, \& {titi}}]{2011TPyx}
{Chesneau}, O., {Meilland}, A., {Banerjee}, D. P.~K., {et~al.} 2011, \aap,
  accepted

\bibitem[{{Cranmer}(2005)}]{2005ApJ...634..585C}
{Cranmer}, S.~R. 2005, \apj, 634, 585

\bibitem[{{de Marco}(2009)}]{2009PASP..121..316D}
{de Marco}, O. 2009, \pasp, 121, 316

\bibitem[{{Delaa} {et~al.}(2011){Delaa}, {Stee}, {Meilland}, {Zorec},
  {Mourard}, {B{\'e}rio}, {Bonneau}, {Chesneau}, {Clausse}, {Cruzalebes},
  {Perraut}, {Marcotto}, {Roussel}, {Spang}, {McAlister}, {Ten Brummelaar},
  {Sturmann}, {Sturmann}, {Turner}, {Farrington}, \&
  {Goldfinger}}]{2011A&A...529A..87D}
{Delaa}, O., {Stee}, P., {Meilland}, A., {et~al.} 2011, \aap, 529, A87+

\bibitem[{{Domiciano de Souza} {et~al.}(2007){Domiciano de Souza}, {Driebe},
  {Chesneau}, {Hofmann}, {Kraus}, {Miroshnichenko}, {Ohnaka}, {Petrov},
  {Preisbisch}, {Stee}, {Weigelt}, {Lisi}, {Malbet}, \&
  {Richichi}}]{2007A&A...464...81D}
{Domiciano de Souza}, A., {Driebe}, T., {Chesneau}, O., {et~al.} 2007, \aap,
  464, 81

\bibitem[{{Domiciano de Souza} {et~al.}(2003){Domiciano de Souza}, {Kervella},
  {Jankov}, {Abe}, {Vakili}, {di Folco}, \& {Paresce}}]{2003A&A...407L..47D}
{Domiciano de Souza}, A., {Kervella}, P., {Jankov}, S., {et~al.} 2003, \aap,
  407, L47

\bibitem[{{Dunham}(1974)}]{Dunham1974}
{Dunham}, D.~W. 1974, in Occultation Newsletter, ed. I.~O.~T. Association,
  Vol.~1, 4

\bibitem[{{Fabregat} {et~al.}(2000){Fabregat}, {Reig}, \&
  {Otero}}]{2000IAUC.7461....1F}
{Fabregat}, J., {Reig}, P., \& {Otero}, S. 2000, \iaucirc, 7461, 1

\bibitem[{{Fr{\'e}mat} {et~al.}(2005){Fr{\'e}mat}, {Zorec}, {Hubert}, \&
  {Floquet}}]{2005A&A...440..305F}
{Fr{\'e}mat}, Y., {Zorec}, J., {Hubert}, A.-M., \& {Floquet}, M. 2005, \aap,
  440, 305

\bibitem[{{Hanbury Brown} {et~al.}(1974){Hanbury Brown}, {Davis}, \&
  {Allen}}]{1974MNRAS.167..121H}
{Hanbury Brown}, R., {Davis}, J., \& {Allen}, L.~R. 1974, \mnras, 167, 121

\bibitem[{{Hartkopf} {et~al.}(1996){Hartkopf}, {Mason}, \&
  {McAlister}}]{1996AJ....111..370H}
{Hartkopf}, W.~I., {Mason}, B.~D., \& {McAlister}, H.~A. 1996, \aj, 111, 370

\bibitem[{{Innes}(1901)}]{1901MNRAS..61..358I}
{Innes}, R.~T.~A. 1901, \mnras, 61, 358

\bibitem[{{Kanaan} {et~al.}(2008){Kanaan}, {Meilland}, {Stee}, {Zorec},
  {Domiciano de Souza}, {Fr{\'e}mat}, \& {Briot}}]{2008A&A...486..785K}
{Kanaan}, S., {Meilland}, A., {Stee}, P., {et~al.} 2008, \aap, 486, 785

\bibitem[{{Kervella} \& {Domiciano de Souza}(2006)}]{2006A&A...453.1059K}
{Kervella}, P. \& {Domiciano de Souza}, A. 2006, \aap, 453, 1059

\bibitem[{{Kervella} {et~al.}(2008){Kervella}, {Domiciano de Souza}, \&
  {Bendjoya}}]{2008A&A...484L..13K}
{Kervella}, P., {Domiciano de Souza}, A., \& {Bendjoya}, P. 2008, \aap, 484,
  L13

\bibitem[{{Koubsk{\'y}} {et~al.}(2006){Koubsk{\'y}}, {Harmanec}, {Yang},
  {Netolick{\'y}}, {{\v S}koda}, {{\v S}lechta}, \& {Kor{\v
  c}{\'a}kov{\'a}}}]{2006A&A...459..849K}
{Koubsk{\'y}}, P., {Harmanec}, P., {Yang}, S., {et~al.} 2006, \aap, 459, 849

\bibitem[{{Kraus} {et~al.}(2011){Kraus}, {Monnier}, {Che}, {Schaefer},
  {Touhami}, {Gies}, {Aufdenberg}, {Baron}, {Thureau}, {Brummelaar},
  {McAlister}, {Turner}, {Sturmann}, \& {Sturmann}}]{2011arXiv1109.3447K}
{Kraus}, S., {Monnier}, J.~D., {Che}, X., {et~al.} 2011, ArXiv e-prints

\bibitem[{{Labeyrie} {et~al.}(1974){Labeyrie}, {Bonneau}, {Stachnik}, \&
  {Gezari}}]{1974ApJ...194L.147L}
{Labeyrie}, A., {Bonneau}, D., {Stachnik}, R.~V., \& {Gezari}, D.~Y. 1974,
  \apjl, 194, L147+

\bibitem[{{Li} {et~al.}(2008){Li}, {Cassinelli}, {Brown}, {Waldron}, \&
  {Miller}}]{2008ApJ...672.1174L}
{Li}, Q., {Cassinelli}, J.~P., {Brown}, J.~C., {Waldron}, W.~L., \& {Miller},
  N.~A. 2008, \apj, 672, 1174

\bibitem[{{Meilland} {et~al.}(2011{\natexlab{a}}){Meilland}, {Delaa}, {Stee},
  {Kanaan}, {Millour}, {Mourard}, {Bonneau}, {Petrov}, {Nardetto}, {Marcotto},
  {Roussel}, {Clausse}, {Perraut}, {McAlister}, {Ten Brummelaar}, {Sturmann},
  {Sturmann}, {Turner}, {Ridgway}, {Farrington}, \&
  {Goldfinger}}]{2011A&A...532A..80M}
{Meilland}, A., {Delaa}, O., {Stee}, P., {et~al.} 2011{\natexlab{a}}, \aap,
  532, A80+

\bibitem[{{Meilland} {et~al.}(2010){Meilland}, {Kanaan}, {Borges Fernandes},
  {Chesneau}, {Millour}, {Stee}, \& {Lopez}}]{2010A&A...512A..73M}
{Meilland}, A., {Kanaan}, S., {Borges Fernandes}, M., {et~al.} 2010, \aap, 512,
  A73+

\bibitem[{{Meilland} {et~al.}(2011{\natexlab{b}}){Meilland}, {Millour}, {Stee},
  {Chesneau}, {Borges Fernandes}, , {Millour}, \& {Lopez}}]{2011Be}
{Meilland}, A., {Millour}, F., {Stee}, P., {et~al.} 2011{\natexlab{b}}, \aap,
  submitted

\bibitem[{{Meilland} {et~al.}(2007{\natexlab{a}}){Meilland}, {Millour}, {Stee},
  {Domiciano de Souza}, {Petrov}, {Mourard}, {Jankov}, {Robbe-Dubois}, {Spang},
  {Aristidi}, {Antonelli}, {Beckmann}, {Bresson}, {Chelli}, {Dugu{\'e}},
  {Duvert}, {Gennari}, {Gl{\"u}ck}, {Kern}, {Lagarde}, {Le Coarer}, {Lisi},
  {Malbet}, {Perraut}, {Puget}, {Rantakyr{\"o}}, {Roussel}, {Tatulli},
  {Weigelt}, {Zins}, {Accardo}, {Acke}, {Agabi}, {Altariba}, {Arezki}, {Baffa},
  {Behrend}, {Bl{\"o}cker}, {Bonhomme}, {Busoni}, {Cassaing}, {Clausse},
  {Colin}, {Connot}, {Delboulb{\'e}}, {Driebe}, {Feautrier}, {Ferruzzi},
  {Forveille}, {Fossat}, {Foy}, {Fraix-Burnet}, {Gallardo}, {Giani}, {Gil},
  {Glentzlin}, {Heiden}, {Heininger}, {Hernandez Utrera}, {Hofmann}, {Kamm},
  {Kiekebusch}, {Kraus}, {Le Contel}, {Le Contel}, {Lesourd}, {Lopez}, {Lopez},
  {Magnard}, {Marconi}, {Mars}, {Martinot-Lagarde}, {Mathias}, {M{\`e}ge},
  {Monin}, {Mouillet}, {Nussbaum}, {Ohnaka}, {Pacheco}, {Perrier}, {Rabbia},
  {Rebattu}, {Reynaud}, {Richichi}, {Robini}, {Sacchettini}, {Schertl},
  {Sch{\"o}ller}, {Solscheid}, {Stefanini}, {Tallon}, {Tallon-Bosc}, {Tasso},
  {Testi}, {Vakili}, {von der L{\"u}he}, {Valtier}, {Vannier}, \&
  {Ventura}}]{2007A&A...464...73M}
{Meilland}, A., {Millour}, F., {Stee}, P., {et~al.} 2007{\natexlab{a}}, \aap,
  464, 73

\bibitem[{{Meilland} {et~al.}(2008){Meilland}, {Millour}, {Stee}, {Spang},
  {Petrov}, {Bonneau}, {Perraut}, \& {Massi}}]{2008A&A...488L..67M}
{Meilland}, A., {Millour}, F., {Stee}, P., {et~al.} 2008, \aap, 488, L67

\bibitem[{{Meilland} {et~al.}(2007{\natexlab{b}}){Meilland}, {Stee}, {Vannier},
  {Millour}, {Domiciano de Souza}, {Malbet}, {Martayan}, {Paresce}, {Petrov},
  {Richichi}, \& {Spang}}]{2007A&A...464...59M}
{Meilland}, A., {Stee}, P., {Vannier}, M., {et~al.} 2007{\natexlab{b}}, \aap,
  464, 59

\bibitem[{{Millour} {et~al.}(2009){Millour}, {Chesneau}, {Borges Fernandes},
  {Meilland}, {Mars}, {Benoist}, {Thi{\'e}baut}, {Stee}, {Hofmann}, {Baron},
  {Young}, {Bendjoya}, {Carciofi}, {Domiciano de Souza}, {Driebe}, {Jankov},
  {Kervella}, {Petrov}, {Robbe-Dubois}, {Vakili}, {Waters}, \&
  {Weigelt}}]{2009A&A...507..317M}
{Millour}, F., {Chesneau}, O., {Borges Fernandes}, M., {et~al.} 2009, \aap,
  507, 317

\bibitem[{{Millour} {et~al.}(2011){Millour}, {Meilland}, {Chesneau}, {Stee},
  {Kanaan}, {Petrov}, {Mourard}, \& {Kraus}}]{2011A&A...526A.107M}
{Millour}, F., {Meilland}, A., {Chesneau}, O., {et~al.} 2011, \aap, 526, A107+

\bibitem[{{Miroshnichenko}(2007)}]{2007ApJ...667..497M}
{Miroshnichenko}, A.~S. 2007, \apj, 667, 497

\bibitem[{{Miroshnichenko} {et~al.}(2003){Miroshnichenko}, {Bjorkman},
  {Morrison}, {Wisniewski}, {Manset}, {Levato}, {Grosso}, {Pollmann}, {Buil},
  \& {Knauth}}]{2003A&A...408..305M}
{Miroshnichenko}, A.~S., {Bjorkman}, K.~S., {Morrison}, N.~D., {et~al.} 2003,
  \aap, 408, 305

\bibitem[{{Miroshnichenko} {et~al.}(2001){Miroshnichenko}, {Fabregat},
  {Bjorkman}, {Knauth}, {Morrison}, {Tarasov}, {Reig}, {Negueruela}, \&
  {Blay}}]{2001A&A...377..485M}
{Miroshnichenko}, A.~S., {Fabregat}, J., {Bjorkman}, K.~S., {et~al.} 2001,
  \aap, 377, 485

\bibitem[{{Moreno} {et~al.}(2011){Moreno}, {Koenigsberger}, \&
  {Harrington}}]{2011A&A...528A..48M}
{Moreno}, E., {Koenigsberger}, G., \& {Harrington}, D.~M. 2011, \aap, 528, A48+

\bibitem[{{Mourard} {et~al.}(2009){Mourard}, {Clausse}, {Marcotto}, {Perraut},
  {Tallon-Bosc}, {B{\'e}rio}, {Blazit}, {Bonneau}, {Bosio}, {Bresson},
  {Chesneau}, {Delaa}, {H{\'e}nault}, {Hughes}, {Lagarde}, {Merlin}, {Roussel},
  {Spang}, {Stee}, {Tallon}, {Antonelli}, {Foy}, {Kervella}, {Petrov},
  {Thiebaut}, {Vakili}, {McAlister}, {ten Brummelaar}, {Sturmann}, {Sturmann},
  {Turner}, {Farrington}, \& {Goldfinger}}]{2009A&A...508.1073M}
{Mourard}, D., {Clausse}, J.~M., {Marcotto}, A., {et~al.} 2009, \aap, 508, 1073

\bibitem[{{Netolick{\'y}} {et~al.}(2009){Netolick{\'y}}, {Bonneau}, {Chesneau},
  {Harmanec}, {Koubsk{\'y}}, {Mourard}, \& {Stee}}]{2009A&A...499..827N}
{Netolick{\'y}}, M., {Bonneau}, D., {Chesneau}, O., {et~al.} 2009, \aap, 499,
  827

\bibitem[{{Okazaki}(1997)}]{1997A&A...318..548O}
{Okazaki}, A.~T. 1997, \aap, 318, 548

\bibitem[{{Otero} {et~al.}(2001){Otero}, {Fraser}, \&
  {Lloyd}}]{2001IBVS.5026....1O}
{Otero}, S., {Fraser}, B., \& {Lloyd}, C. 2001, Information Bulletin on
  Variable Stars, 5026, 1

\bibitem[{{Petrov} {et~al.}(2007){Petrov}, {Malbet}, {Weigelt}, {Antonelli},
  {Beckmann}, {Bresson}, {Chelli}, {Dugu{\'e}}, {Duvert}, {Gennari},
  {Gl{\"u}ck}, {Kern}, {Lagarde}, {Le Coarer}, {Lisi}, {Millour}, {Perraut},
  {Puget}, {Rantakyr{\"o}}, {Robbe-Dubois}, {Roussel}, {Salinari}, {Tatulli},
  {Zins}, {Accardo}, {Acke}, {Agabi}, {Altariba}, {Arezki}, {Aristidi},
  {Baffa}, {Behrend}, {Bl{\"o}cker}, {Bonhomme}, {Busoni}, {Cassaing},
  {Clausse}, {Colin}, {Connot}, {Delboulb{\'e}}, {Domiciano de Souza},
  {Driebe}, {Feautrier}, {Ferruzzi}, {Forveille}, {Fossat}, {Foy},
  {Fraix-Burnet}, {Gallardo}, {Giani}, {Gil}, {Glentzlin}, {Heiden},
  {Heininger}, {Hernandez Utrera}, {Hofmann}, {Kamm}, {Kiekebusch}, {Kraus},
  {Le Contel}, {Le Contel}, {Lesourd}, {Lopez}, {Lopez}, {Magnard}, {Marconi},
  {Mars}, {Martinot-Lagarde}, {Mathias}, {M{\`e}ge}, {Monin}, {Mouillet},
  {Mourard}, {Nussbaum}, {Ohnaka}, {Pacheco}, {Perrier}, {Rabbia}, {Rebattu},
  {Reynaud}, {Richichi}, {Robini}, {Sacchettini}, {Schertl}, {Sch{\"o}ller},
  {Solscheid}, {Spang}, {Stee}, {Stefanini}, {Tallon}, {Tallon-Bosc}, {Tasso},
  {Testi}, {Vakili}, {von der L{\"u}he}, {Valtier}, {Vannier}, \&
  {Ventura}}]{2007A&A...464....1P}
{Petrov}, R.~G., {Malbet}, F., {Weigelt}, G., {et~al.} 2007, \aap, 464, 1

\bibitem[{{Plets} {et~al.}(1995){Plets}, {Waelkens}, \&
  {Trams}}]{1995A&A...293..363P}
{Plets}, H., {Waelkens}, C., \& {Trams}, N.~R. 1995, \aap, 293, 363

\bibitem[{{Rivinius} {et~al.}(1998){Rivinius}, {Baade}, {Stefl}, {Stahl},
  {Wolf}, \& {Kaufer}}]{1998A&A...336..177R}
{Rivinius}, T., {Baade}, D., {Stefl}, S., {et~al.} 1998, \aap, 336, 177

\bibitem[{{Tango} {et~al.}(2009){Tango}, {Davis}, {Jacob}, {Mendez}, {North},
  {O'Byrne}, {Seneta}, \& {Tuthill}}]{2009MNRAS.396..842T}
{Tango}, W.~J., {Davis}, J., {Jacob}, A.~P., {et~al.} 2009, \mnras, 396, 842

\bibitem[{{Tycner} {et~al.}(2011){Tycner}, {Ames}, {Zavala}, {Hummel},
  {Benson}, \& {Hutter}}]{2011ApJ...729L...5T}
{Tycner}, C., {Ames}, A., {Zavala}, R.~T., {et~al.} 2011, \apjl, 729, L5+

\bibitem[{{Vinicius} {et~al.}(2006){Vinicius}, {Zorec}, {Leister}, \&
  {Levenhagen}}]{2006A&A...446..643V}
{Vinicius}, M.~M.~F., {Zorec}, J., {Leister}, N.~V., \& {Levenhagen}, R.~S.
  2006, \aap, 446, 643

\end{thebibliography}

\end{document}